# A Novel Model for Integration of Information Security Management against Replication Attack Based on Biological Structures of the Body

**Amir Hossein Bodaghi**

Ministry of Information and Communication Technology, Organization of Information Technology, Tehran, Iran

Tel:(+98)9126230952 E-mail: itworker2012@gmail.com

**Abstract**

By developing communications and increase of access points, computer networks have been vulnerable considerably against wide range of information attacks, specially new and complicated attacks. Every day, replication attacks attack millions of network and mobile users. Increase in amount of replication attack may be a potential danger for income of SMS or network and causes losing customers of these services provider. Humans or software can be used to encounter these replication attacks. It is obvious that lonely absolute use of each method will not result in a proper answer to encounter replication attack`s problem. Since replication attack is one of the important problems of information protection and security in organizations for computer and mobile phone users, while reviewing types of replication attacks and methods of encountering, this paper uses similarities between pathologies in body and invader factors in replication attacks, a model is provided based on biological simulation methods existing in body`s adapted immune system to encounter these threats.

## 1. Introduction

Today, information considers as an essential good for organizations, and it`s vulnerability is an essential concern specially for big organizations equipped with computer networks. Networks and information have been increasingly vulnerable against intrusion attacks, especially new direct and remote controlled methods. Based on researches of IDC (international Data Corporation), (short transaction service) SMS almost forms almost 25% of mobile phone service provider`s earning. Based on a new definition, actions which put integrity, privately or accessibility of resources in danger are called “intrusion” [1]. Based on this definition, there are three main threats for information security [2,3]:

- Threat of unauthorized access to information
- Threat of destroying information integrity
- Threat of unavailability of vital resources and information in case of necessity

Replication attack equals massive useless networks. Against traditional useless mails which were often expensive and time consuming, replication attack may impose high expenses to internet service providers and receivers. It is noteworthy to say that it is almost impossible to indentify all replication attacks based on the currently used rules in filtering systems. Solutions that should be considered to encounter this problem are divided into legitimate and technical solutions. In this case, while some countries have applied legal and lawful solutions, necessity of technical methods still have special importance, because replication attacks can be sent virtually from everywhere, and tracing their senders is not often easily possible. Recently, artificial immune system have been proposed as a soft computational pattern [5] which consists of comprehensive ability of identifying immune system [6,7].

Innate and Adaptive artificial immune system (IA-AIS) [8] is studied in this paper which is created by attaching negative selection algorithm (Ji, Dasgupta) [9] and selecting colonial. In this model, macrophages are used along with (helper and controller) T and B cells. Motive of using this combinations and reaction between them is protected for a biological model designer by Stepney et al. Recently, innate immune system at artificial immune system has attracted attention. Green smith et al proposedDandritic cells algoriths (DCA), while Tedesco et al provided a method to attach Dandritic cells and T cells. Also Sarafijanoic and Le Boudec andDasgupta et al , used innate systems. In some algorithms, like Timmis&Freitas[10] and Kim &Bently work, provided colonial dynamic selection algorithm.

**2. Body`s immune system versus artificial immune system**

Body`s immune system is capable of detecting self and non self cells while artificial immune systems are capable of identifying legitimate self (non replication attack) networks and illegal non self networks (replication attacks). Despite that, body`s immune system has preferences in relation with anti replication attack immune system. Self cells in a biologic system do not change if no problems occur at immune system, because proteins which use to differentiate self cells remain constant during their life time. Unfortunately, anti replication attack immune system has the same problem as protective – security system of computers [11].

For better comprehension of artificial immune system, we need to comprehend body`s immune system because it`s the basis of artificial designed system. So it is necessary to simply and briefly review operation of body`s immune system [12]. Identifying and differentiating between self and potentially dangerous and none self-harmful cells is the first and the most important duty of an immune system. Pathogens are those none self-harmful elements. Since adapted immune system is the basis of imitating anti replication attack immune system, so it will be explained with more details.

Adapted immune system mainly consists of lymphocytes which are considered type of white globules that should identify and destroy pathogens. Lymphocytes identify pathogens by connecting their selves to them. To produce Lymphocytes, body uses a library of different genes that randomly join each other to produce different anti bodies. This library has different gene parts which use to produce different identifiers to identify different pathogens. To create a storage or density of Lymphocyte in each time, elements of this genius library combine randomly and create different types of receivers (figure 1).

Digital anti bodies for artificial immune system applied in replication attack are also created by a similar method and by using a library. But in this type of immune system, instead of gene`s different parts, library has different parts of adaptability patterns based on texts, which are explicitly that sentences or expressions.From gradual changes point of view, an adapted immune system is important because it is able to identify new pathogen factors which try to hide from immune system. Main benefit of having two innate and adapted immune systems is related to reactions between them, this provides possibility of fast and correct answer toward pathogens. Innate immune system creates signaling proteins called Cytoxine and causes inflammation and activation of adapted immune system. On the other hand, adapted system uses cells of innate immune system to destroy pathogens. Colony selection theory (CST) provides a model to establish and maintain “immune memory”. The basis is that, after destroying invader pathogens, some of created colons transform into immune system and cause creating a structure which enables body to act faster when reencountering above or the same pathogens. One main idea in CST is creating a distinction between self and non selves; so that, it is as an ability of system to encounter non selves and act against them, which are those exterior and harmful factors (pathogens) and no act against selves and non-harmful interior factors .

**3. Proposed Model Elements**

In our model for categorizing transactions, a system has proposed to arrange transaction to classify transactions, in which user`s received feedback, as a second signal, results in creating an active cell. It provided an adaptive error detection (AED) system. In the model, a transaction is shown as a microorganism in which transaction head and body that always consist of content, is considered equal to anti genes. Sender`s address will be considered the same as molecular pattern (figure 2). This definition is related to type of application and should be done in a way that, that type of input showing none self (transaction sender`s address) considers as molecular pattern, while other properties (transaction head and body) which are dependent to each other and need more analyze, are used to create anti genes. This is because molecular patterns detect by macrophage while T and B cells are responsible for identify anti genes. About replication attack massages, this definition of microorganisms originates from the point that all received transactions after system training, which their sender`s address is presented as replication attack during levels of system training, are considered as replication attack.

To produce microorganisms, characters are coded using 6 bits coding system which is exclusively designed to encounter this problem. As Freitas and Timmis have mentioned in their article, in this coding method, characters with the same appearance (like 0 and O, 3 and E) are different just in one or two codes.

So, sequence in content, like transaction content, show as an anti-gene in which each word equals a peptide in an anti-gene. To measure affinity value relative number of continues bits are used in two sequences. If activity threshold is defined accurately, using this method it is possible that TEST receiver cell join TEST anti gene. Often in this case, alternate spelling does not prevent from identifying applied pattern. As shown in figure 3, by arranging anti gens and molecular patterns applied of information used to train system, it prepares, this is done by characterizing content of replication attacks and legitimate transactions about self or non-self-mails.

Macrophage create using molecular pattern derived of non self inputs and do not exist in self patterns. On the other hand, controller T cells, create using their random selected anti genes that use to code their receivers. Creating helper T and B cells also follows a similar procedure by using non self anti genes. After this level, is negative selection level, in which remaining lymphocytes go to next level by removing identified self anti genes. As shown in figure 4, in categorization, micro organisms first show to macrophage group that analyzes their molecular pattern. If at least a macrophage activated, micro organism categorizes as non self and adapted immune system stimulates by creating B cells and helper T cells, specially related anti gene to that micro organism. Otherwise, procedure continues by showing anti genes to set of B cells. If no B cell connects to this anti gene, micro organism will be considered as self and analyze ends. On the other hand, if at least o B cell stimulates, anti gene will be stimulated by B lymphocyte, introduces to T cells group.

Micro organism categorize as self and non self cells, based on T cells reaction. To activation, stimulated B cell should reactwith assistant and controller T cell. Before this level, B cell passes a procedure through producing sequences with peptide (words) that results in cell stimulation. In proposed model, this method is used to justify manner of showing derived peptides of anti genes connected to MHC molecule. During stimulation, B cell is stimulated and receives stimulator and restraining signals that creates by assistant and controller T cells. Stimulator signals to identify non self, creates by assistant T cells to stimulate creating response in system.

On the other hand, restraining signals are related to transaction identification as self that result in restraining stimulated cell. Final result is related to value of each signal whether stimulates B cell more or restrain it. B cell stimulates only when sum of timesthat all helper T cells are stimulated are more than timesof stimulating controller T cell. This reaction is a mechanism to control simulated B cells with self cells and able to identify phrases that exist in legitimatetransactions and may passed through negative selection level.

When a cell establishes, a survival time defines that is an integer for showing countdown of reaching to death of cell. This value decreases after each time of encountering anti gene, till reaches zero. A special reward of cell is used to stimulate competition between cells to identify anti gene. So, for stimulated cells, their life time increase using reward, while in case of no stimulation, they will be replaced by new cells. A similar method is proposed by Cutello et al . In this method, cell removes by reaching to a determined value. When cell clones, it`s lifetime regulates again. During micro organism categorization, system categorizes by decreasing lifetime. So cells with zero life time get removed and new cell will be create using new molecular pattern. To ensure that group size still constant, there may be need to add new cells and remove old cells with ended or ending lifetime. Pattern of keeping the size constant, is a way of system simplification and allows us for easily existence of suitable cells during primary searching level. Procedure of correcting a categorizer with correcting wrong positive and wrong negative cases has been shown in table 1. In each case it is necessary to remove saved pattern and produce proper cells. Amount of user relation should be explicit here. Categorizer amplifier level (about correct function) or its correction (in case of wrong function) also may be foresighted in this section.

Creating a library of the words used in legitimate and replication attacktransaction is one of the necessities in this model (like place of keeping genes in body). Information about contact (telephone no- web address) header information is also used in library to create anti bodies. Then set of Networktransactions enter training system which has been divided into replication attack and none replication attack previously. Lymphocytes create and train using both transactions. When system is trained and used, lymphocytes also train in case of receiving new transactions. Lymphocytes get select, old, and new ones create.

*3.1 Digital library*

Genetic library consists of preselected patterns used to create applied patterns in lymphocytes. Selected patterns in this case can be little patterns merges together and provide main plan which is necessary for lymphocytes and or may be individual patterns used in defining necessary pattern in lymphocyte. For an anti replication attack immune system we can also use little patterns which will help us in finding legitimate or replication attacktransactions, in case of combining together or individually. Main purpose of creating this library is to create a place to produce lymphocytes that help us identifying transactions such as replication attack or legitimate. It should be noted that lymphocytes which do not match any replication attack or legitimate transaction will not be suitable for us since the system cannot use them correctly and properly to decide about identifying transactions. A useful lymphocyte is the one which has been matched with legitimate transactions at least one time and has bigger than zero value of legitimate transactionmatch. More explicitly, each lymphocyte having zero value of legitimate transaction match cannot help in this matter. Next purpose of using library is to continuously create lymphocytes with high adaptability. The more number of lymphocytes adapt with transaction, shows better function of system in categorizing transactions and system will have higher separation capability. Whatever number of lymphocytes with zero match times is less, it is better for system because time of processes and their amount of occupied memory will be saved. These lymphocytes are called high efficiency lymphocytes and it does not seem suitable to remove them totally. lymphocyte`s life time is also important. Of course, the life time is dependant to the time of adapting

*3.2 Weight or value*

Each lymphocyte consists of two types of weight or information value, replication attack match and transaction match. Both of these numbers are set to zero at the beginning. If an identifier adapt with a transaction, amount of transaction match will increase one unit and also if a transaction has identified as replication attack, amount of replication attack match will increase one unit. These numbers will change based on amount of categorized error in primary stages of education, normal function and next relearning.

*3.3 Learning stage*

At primary learning stage, system`s considered transaction are categorized previously. Hence, if the transaction is replication attack, amount of replication attack match increases one unit, and if the transaction is legitimate, the number equals zero.

*3.4 Normal function*

In normal function, system will have no information about input transactions and system will identify it by itself. So, amount of replication attack match will be noticed by identified point of system.

*3.5 Relearning stage*

If we want to learn the system again, it means that the system had mistaken in categorization, and based on these mistakes the system will be learned again. So, new values should be added to system along with relearning the system. Relearning equals repeating normal learning, with this difference which in relearning we should have primary values assigned to mistake learned lymphocytes. Hence, that numbers will change into user defined values.

*3.6 Test stage*

In testing amounts of replication attack or legitimate transaction match, both of them are updated. According to above assumptions, a flowchart of operational stages in investigative anti replication attack program has been designed.

**4. Conclusion**

In this paper, we studied method derived of human immune system to create information security about dangers of replication attack and its threats. This model consists of macrophage, as innate immune system marker; and B and T cells as adaptive immune system. Replication attack identifies based on detecting sender`s address by macrophage and its subject and content identifies by B and T cells. An important characteristic is that in this method, feedback information of user is used. Categorizer is able to update its information and acquire information which has not received during training phase through user feedback.

**References**

[1] S. Forrest, S. A. Hofmeyer, and A.Somayaji, "Computer Immunology," *Communication of The ACM*, 2001

[2] J. Balasubramaniyan, J. Garcia-Fernandez, D. Isakoff, E. Spafford, and D. Zamboni, " An Architecture for Intrusion Detection using Autonomous Agents," *In Proceedings of the Annual Computer SecurityApplications Conference*, Phoenix Arizona , 2006

[3] G. White, E. Fish, and U. Pooch, "Cooperating Security Managers: A Peer-Based Intrusion Detection System*,"IEEE Network*, 2008

[4] Carpinter, J., Hunt, R., Tightening the net: a review of current and next generation REPLICATION ATTACK filtering tools. *Comput.* 2009

[5] De Castro, L.N., Timmis, J. Artificial Immune Systems: A New Computational Intelligence Approach, st ed. *Springer.*2005

[6] Varela, F.J., Coutinho, A., Dupire, B., Vaz, N.M. Cognitive networks:immune, neural andotherwise. In: Perelson, A.S. (Ed.), Theoretical Immunology. Part 6.SFI *Studies in the Sciences ofComplexity. Addison-Wesley*, Boston, 2000

[7] Cohen, I., 6The cognitive principle challenges clonal selection. Immunol.TodayAtkins, S. Size and cost of the problem. *In Proceedings of the Fifty-sixth Internet Engineering Task Force (IETF) Meeting (San Francisco, CA), Replication attackCon Foundation*, 2009

[8] Forrest, S., and Hofmeyr, S.A., and Somayaji, A. Computer immunology. *Communications of the ACM*, 2004

[9] Janeway, C.A., Travers, P., Walport, M., Shlonmchik, M. The Immune System in Health andDisease, *4th ed. Garland Publishing, Oxford.*2006

[10] A.K Abbas, A.H Litchman, Sh. Pilla , Cellular and Molecular Immunology, *Elsevier IndiaPvt Ltd* 2010

[11] Delany, S.J., Cunningham, P., Tsymbal, A., Coyle, L, A case-based technique for tracking concept drift in SPAM filtering. *Knowl.-Based Syst.* 2004

[12] Wang, B., Jones, G.J.F., Pan, W., 633/. Using online linear classifiers to filter SPAM e-mails. *Pattern Anal.*2007.

Table 1. process of refining the categorizer, a: Correcting Wrong Positive, b: Correcting Wrong Negative

| a: Correcting Wrong Positive |
|---|
| Pick up anti-genes and molecular pattern out of non-self-library |
| Pick up macrophage connecting to the molecular pattern |
| Adding up anti-genes and molecular pattern into self-library |
| Implementing negative selection over help T and B cells |
| Creating controller T cells using anti-gene |

| b: Correcting Wrong Negative |
|---|
| Pick up anti-genes and molecular pattern from self-library |
| Pick up T cells connecting to the anti-genes |
| Adding up anti-genes and molecular pattern to non-self-library |
| Creating macrophage based on molecular pattern |

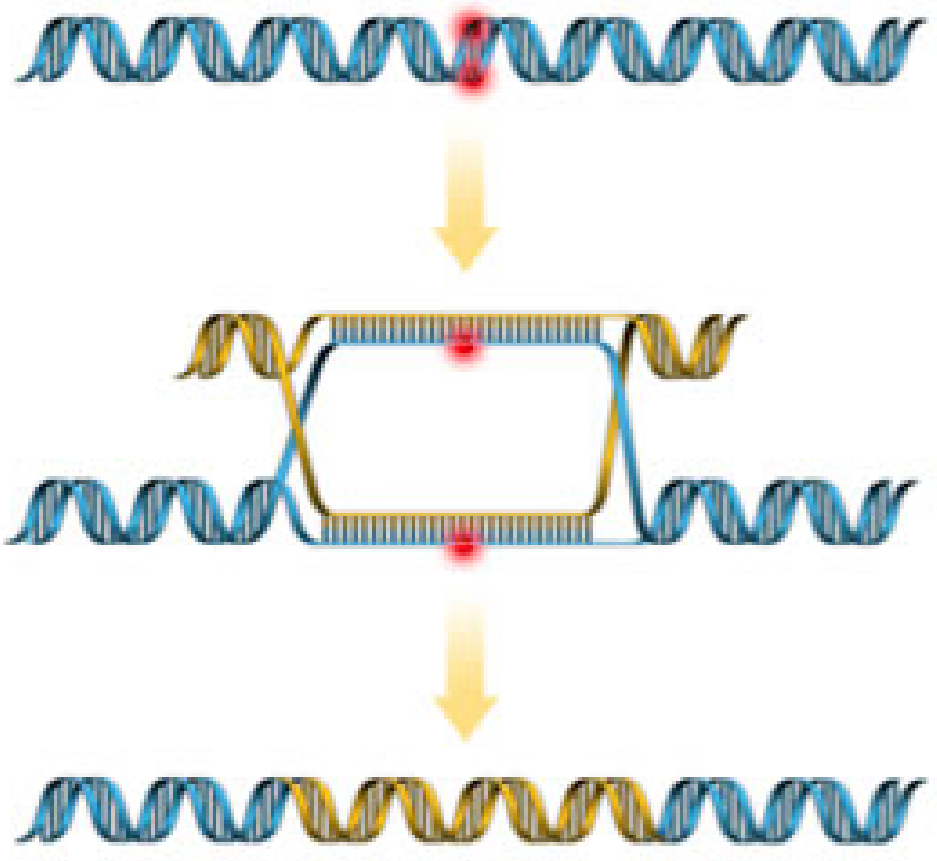

Figure 1. Random connection of gene parts

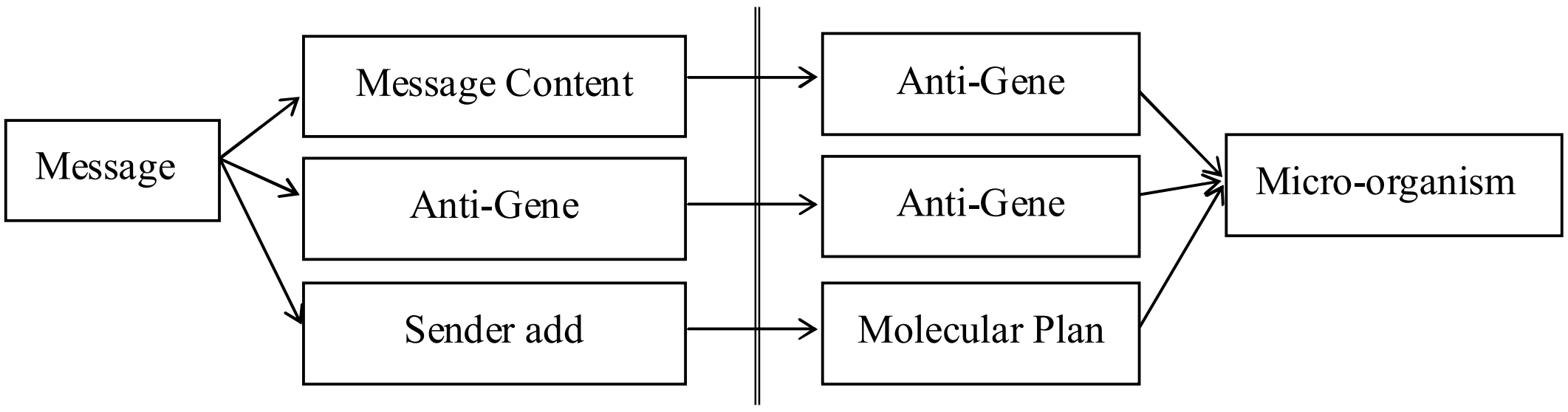

Figure 2. creating micro-organism from network transaction

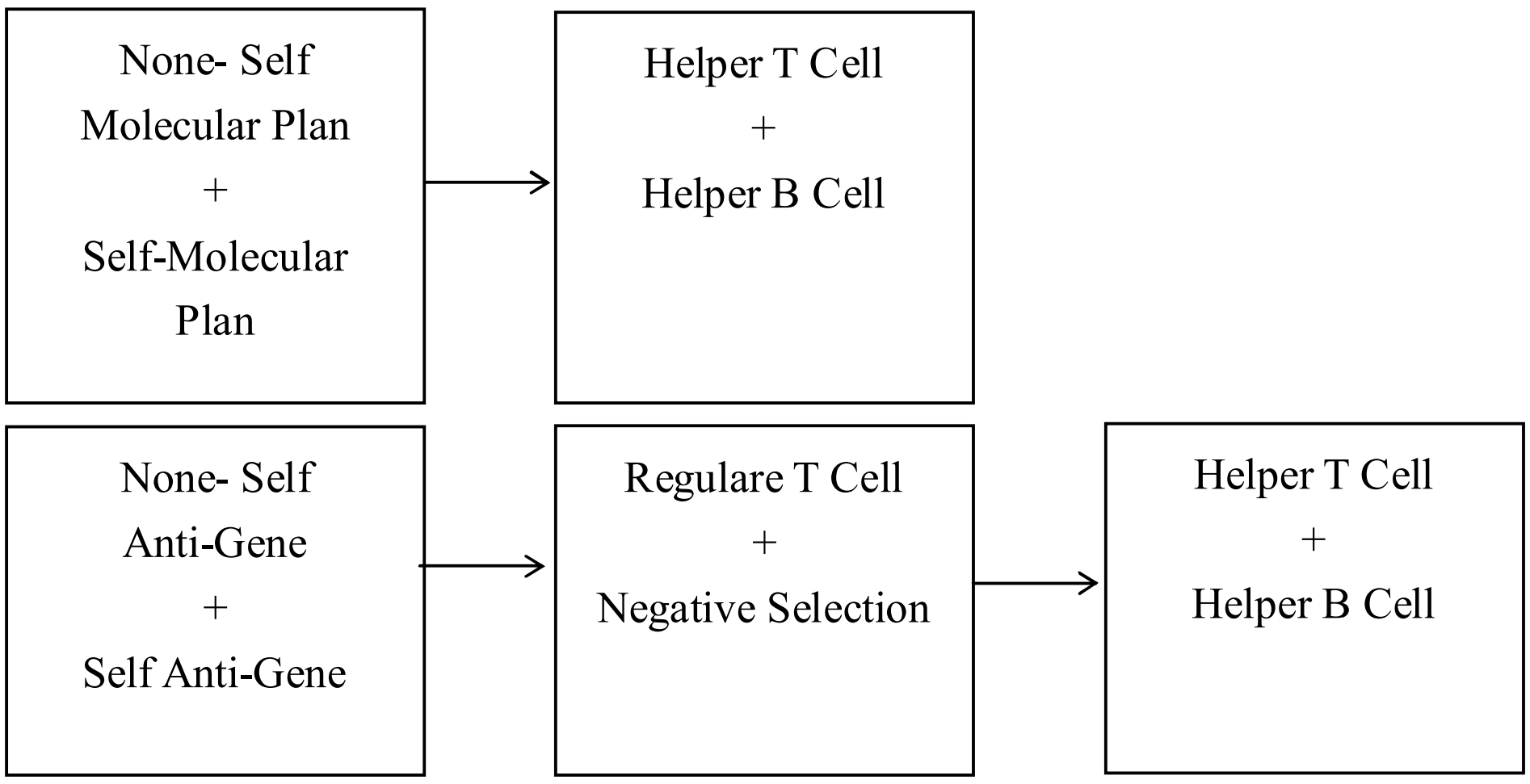

Figure 3: Data collection system along training stage

```
SendingTransaction (S)
  S. Analyzing(Senderaddress)
  If macrophage.activation() then
        S.category = Replicated Transaction
        S. (stimulating adaptive immune system).activate()
 If macrophage.non-activation() then
        S.analyzing(transaction content) by B cell
        If Bcell.immobility()then
            S.category = legitimate
        If B cell.stimulation()   then
            S.analyzing( anti-gen) by T cell
            If Bcell.non-activation()   then   S.category = legitimate
            If Bcell.non-activation()S.clone
                        S.mutation of stimutated cell ()
                        S.category = Replicated Transaction
```

Figure 4: proposed code of the model for transaction categorization